\documentclass{PoS}
\usepackage{color,graphicx}
\usepackage{subfigure}
\usepackage{amsmath}

\title{Partial-Wave Analysis of the Centrally Produced $\pi^+\pi^-$ System in $pp$ Reactions at COMPASS}

\ShortTitle{Partial-Wave Analysis of the Centrally Produced $\pi^+\pi^-$ System in $pp$ Reactions at COMPASS}

\author{Alexander Austregesilo\thanks{The authors acknowledge financial support by the German Bundesministerium f\"ur Bildung und Forschung (BMBF), by the Maier-Leibnitz-Laboratorium der LMU und TU M\"unchen, and by the DFG cluster of excellence 'Origin and Structure of the Universe'.}\\
  Technische Universit\"at M\"unchen\\
  E-mail: \email{alexander.austregesilo@cern.ch}}

\author{\speaker{Tobias Schl\"uter}\\
  Ludwig-Maximilians-Universit\"at M\"unchen\\
  E-mail: \email{tobias.schlueter@physik.uni-muenchen.de}}

\author{for the COMPASS collaboration}

\abstract{COMPASS is a fixed-target experiment at CERN SPS which investigates the structure and spectroscopy of hadrons. During nine weeks in 2008 and 2009, a $190\,\textrm{GeV}/c$ proton beam impinging on a liquid hydrogen target was used in order to study the production of exotic mesons and glueball candidates at central rapidities. As no bias on the production mechanism was introduced by the trigger system, the contribution from diffractive dissociation of the beam proton poses a challenge. We select a centrally produced sample by kinematic cuts and introduce a model to describe the data in terms of partial waves. Preliminary fits are presented, which are consistent with results from previous experiments. Particular attention is paid to the ambiguities in the amplitude analysis of the two-pseudoscalar final state.}

\FullConference{Sixth International Conference on Quarks and Nuclear Physics\\
  April 16-20, 2012\\
  Ecole Polytechnique, Palaiseau,  Paris}

\begin{document}
  
  \section{Introduction}
  \label{sec:mot}

\begin{figure}[ht]
    \begin{minipage}{.55\textwidth}
      \begin{center}
        \vspace{.5cm}
        \includegraphics[width=\textwidth]{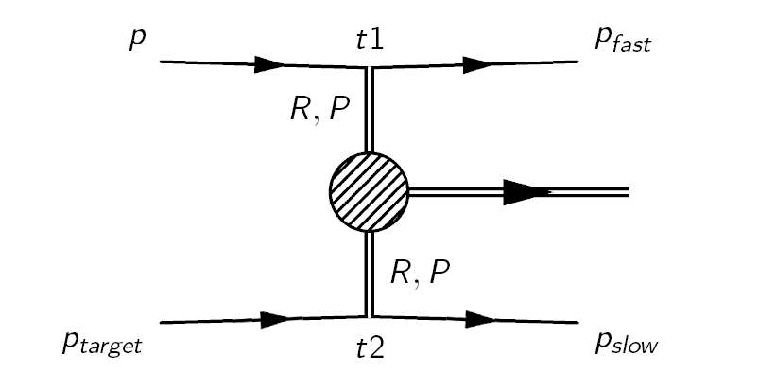}
        \vspace{-.5cm}
        \caption{\em  Central production}
        \label{fig:cp}
      \end{center}
    \end{minipage}
    \begin{minipage}{.44\textwidth}
      \begin{center}
        \includegraphics[width=\textwidth]{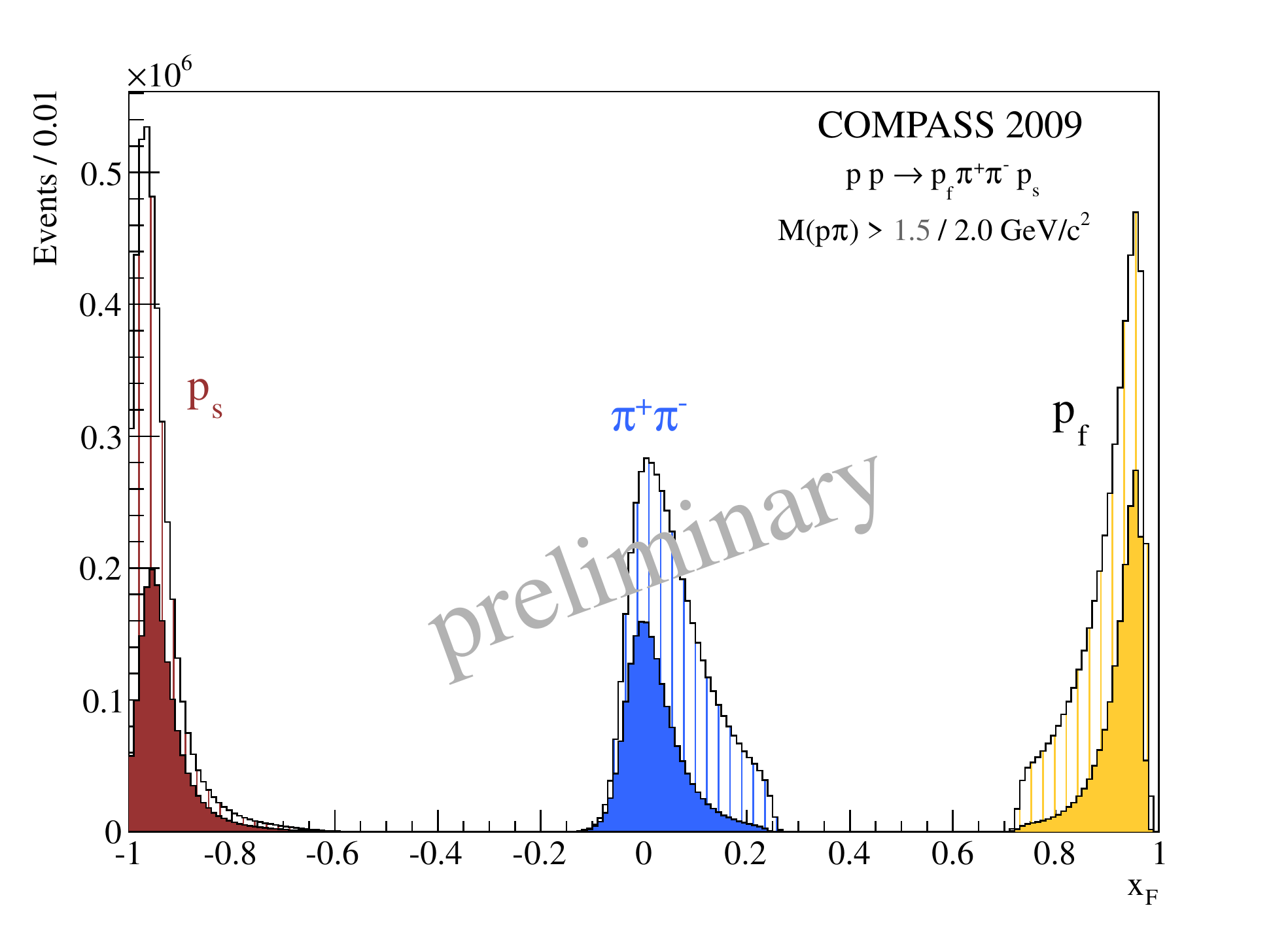}
        \caption[{\em Feynman $x_F$}]{\em  Feynman $x_F$ distributions}% for $p_s$ (red), $p_f$ (yellow) and the $\pi^+\pi^-$ system (blue)}
        \label{fig:xf}
      \end{center}
    \end{minipage}
  \end{figure}

  Quantum Chromodynamics predicts objects composed entirely of valence gluons, so-called {\em glueballs}. Their existence, however, could not be confirmed experimentally to the present day. One of the goals of the COMPASS experiment~\cite{com07} at CERN is to study the existence and signatures of glueballs, in continuation of the efforts that were made at the CERN Omega spectrometer in the late 1990s~\cite{kir97}. Since Pomerons are considered to have no valence quark contribution, Pomeron-Pomeron fusion was proposed to be well suited for the production of glueballs. This process can be realised in a fixed-target experiment by the scattering of a proton beam on a proton target, where a system of particles is produced centrally (cf. Figure~\ref{fig:cp}).

Approximately $30\%$ of the total COMPASS data recorded with a $190\,\textrm{GeV}/c$ proton beam impinging on a liquid hydrogen target was used for the presented analysis. In order to separate the central $\pi^+\pi^-$ system from the fast proton $p_f$, a cut on the invariant mass combinations $M(p\pi) > 1.5\,\mathrm{GeV}/c^2$ was introduced. The effect of this kinematic cut on the Feynman $x_F$ distributions for the fast (yellow) and the slow proton (red) as well as the di-pion system (blue) is shown in Figure~\ref{fig:xf}. The $\pi^+\pi^-$ system  lies within $\left \vert x_F \right \vert \le 0.25$ and can therefore be considered as centrally produced.

\begin{figure}[ht]
    \begin{center}
      \subfigure[{\em $85\,\mathrm{GeV}/c$}~\cite{arm91}]{\includegraphics[width=.2\textwidth]{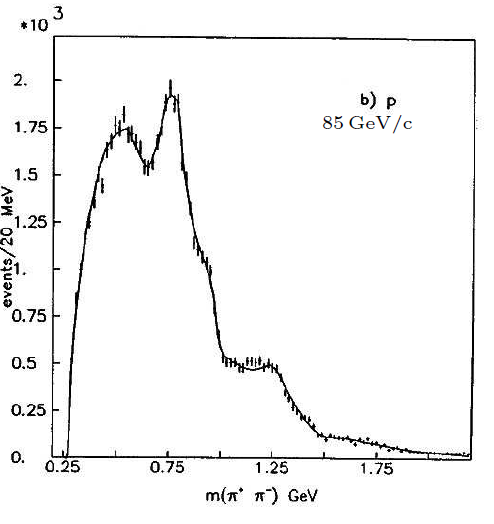}}
      \subfigure[{\em $190\,\mathrm{GeV}/c$}]{\includegraphics[width=.4\textwidth]{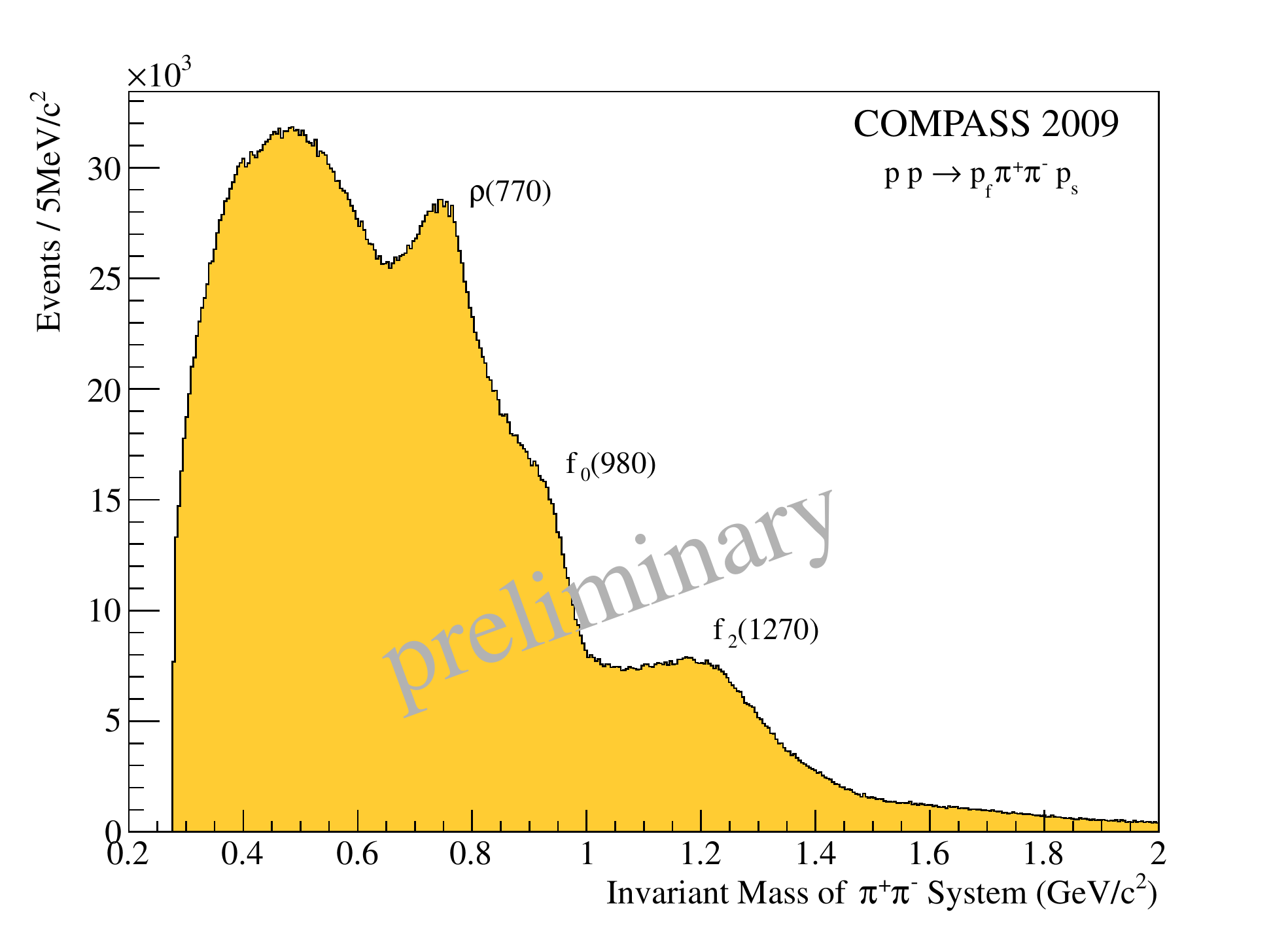}}
      \subfigure[{\em $300\,\mathrm{GeV}/c$}~\cite{arm91}]{\includegraphics[width=.2\textwidth]{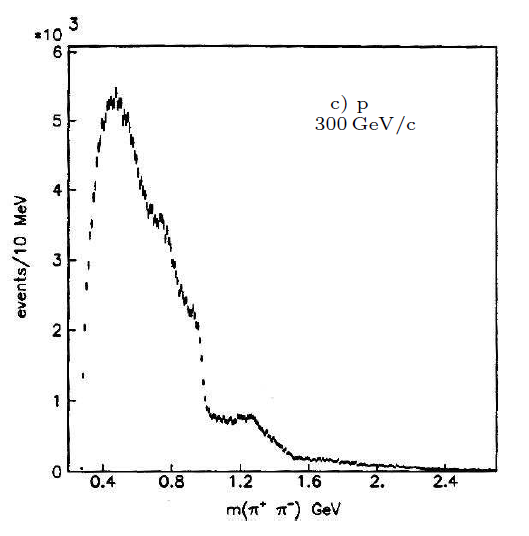}}
    \end{center}
    \caption[{\em $\pi^+\pi^-$ Invariant Mass}]{\em Invariant mass of the $\pi^+\pi^-$ system for three beam energies}
    \label{fig:two}
  \end{figure}

However, central production comprises several production processes, their characteristic signature being the different dependence of the cross section on the centre-of-mass energy $\sqrt s$ of the reaction. While Pomeron-Pomeron scattering should be only weakly dependent on $s$, Pomeron-Reggeon scattering is predicted to scale with $1/\sqrt s$ and Reggeon-Reggeon scattering with $1/s$~\cite{kle07}. This behaviour can be observed experimentally, e.g. by comparing the mass spectra obtained at COMPASS energy to those the Omega spectrometer~\cite{arm91} obtained at different energies (cf. Figure~\ref{fig:two}).

The dominant features of the $\pi^+\pi^-$ invariant mass distribution, the $\rho$(770), the $f_2$(1270) and the sharp drop in intensity in the vicinity of the $f_0$(980), can be observed at all three different centre-of-mass energies. The relative yield of $\rho$-production decreased rapidly with increasing $\sqrt s$, which is explained by a Reggeised pion-pion contribution ($\approx 1/s^2$). On the the other hand, the enhancement at low masses as well as the $f_0$(980) remain practically unchanged, a fact that is characteristic for $s$-independent Pomeron-Pomeron scattering.

  \section{Partial-Wave Analysis}

The partial-wave analysis has been performed assuming that the $\pi^+\pi^-$ system is produced by the collision of two particles emitted by the scattered protons. They are also referred to as exchanged particles, and carry the squared four-momentum transfer $t_1$ from the beam proton and $t_2$ from the target proton, respectively.

We make the strong assumption that $t_1$ only transmits the helicity $\lambda = 0$. In this limit, $t_1$ can be seen as a vacuum-like Pomeron with $J^{PC}=0^{++}$, which is treated like an {\em external particle}. The beam and the fast outgoing proton $p_f$ are irrelevant in this picture. If we accept the space-like Pomeron as the incoming beam, we can construct a Gottfried-Jackson frame~\cite{got64} for the di-pion system. Hence, the exchanged particle $t_1$ in the centre-of-mass frame of the $\pi^+\pi^-$ system defines the $z$-axis for the reaction. We have to emphasise that we fixed the choice to $t_1$ in contrast to previous experiments (e.g. \cite{bar99}) to be able to correct for the different $t$ acceptances of the trigger.

\begin{figure}[ht]
    \subfigure[{\em $\cos(\theta)$}]{\includegraphics[width=.45\textwidth]{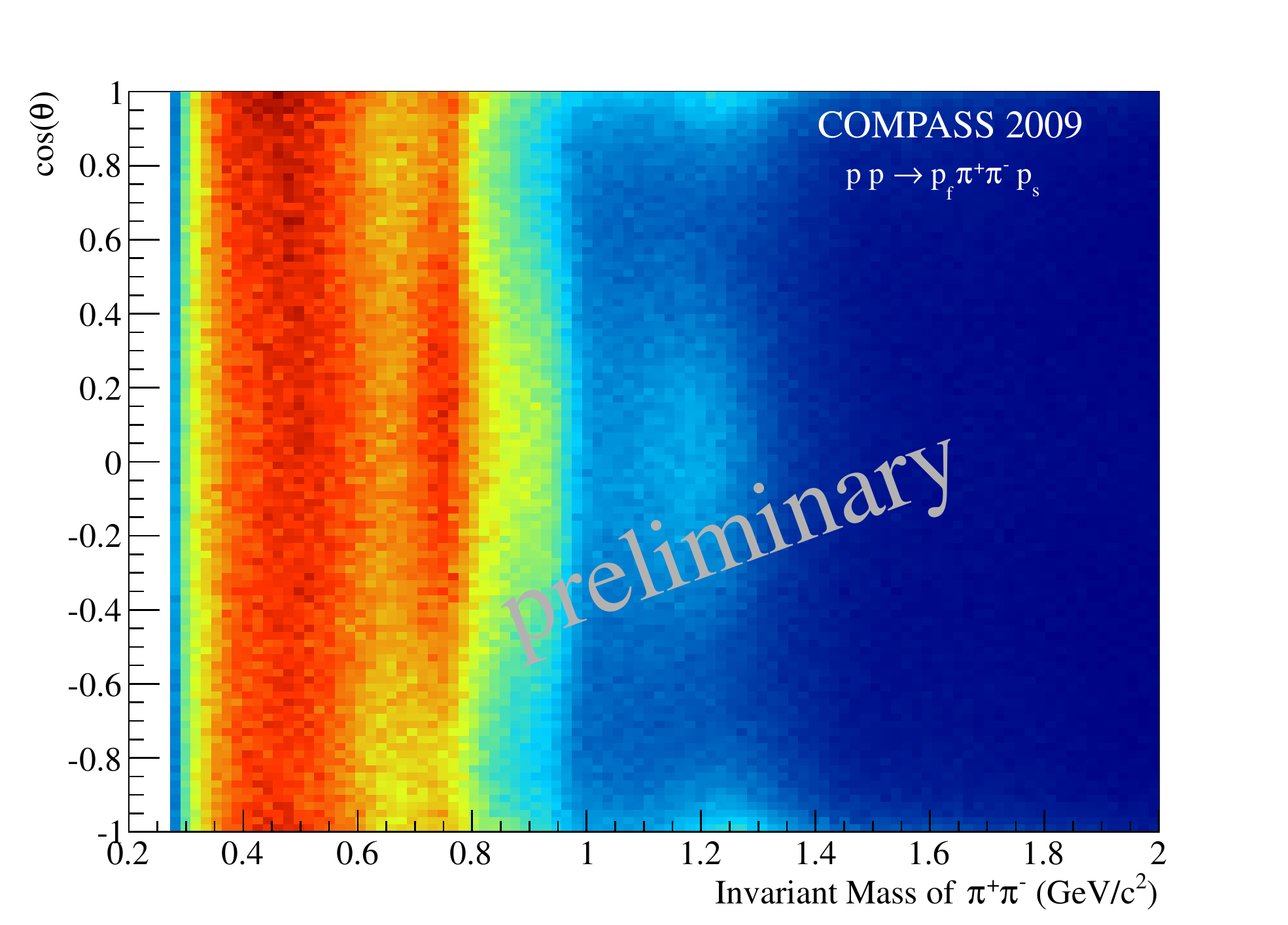}}
    \subfigure[{\em $\phi$}]{\includegraphics[width=.45\textwidth]{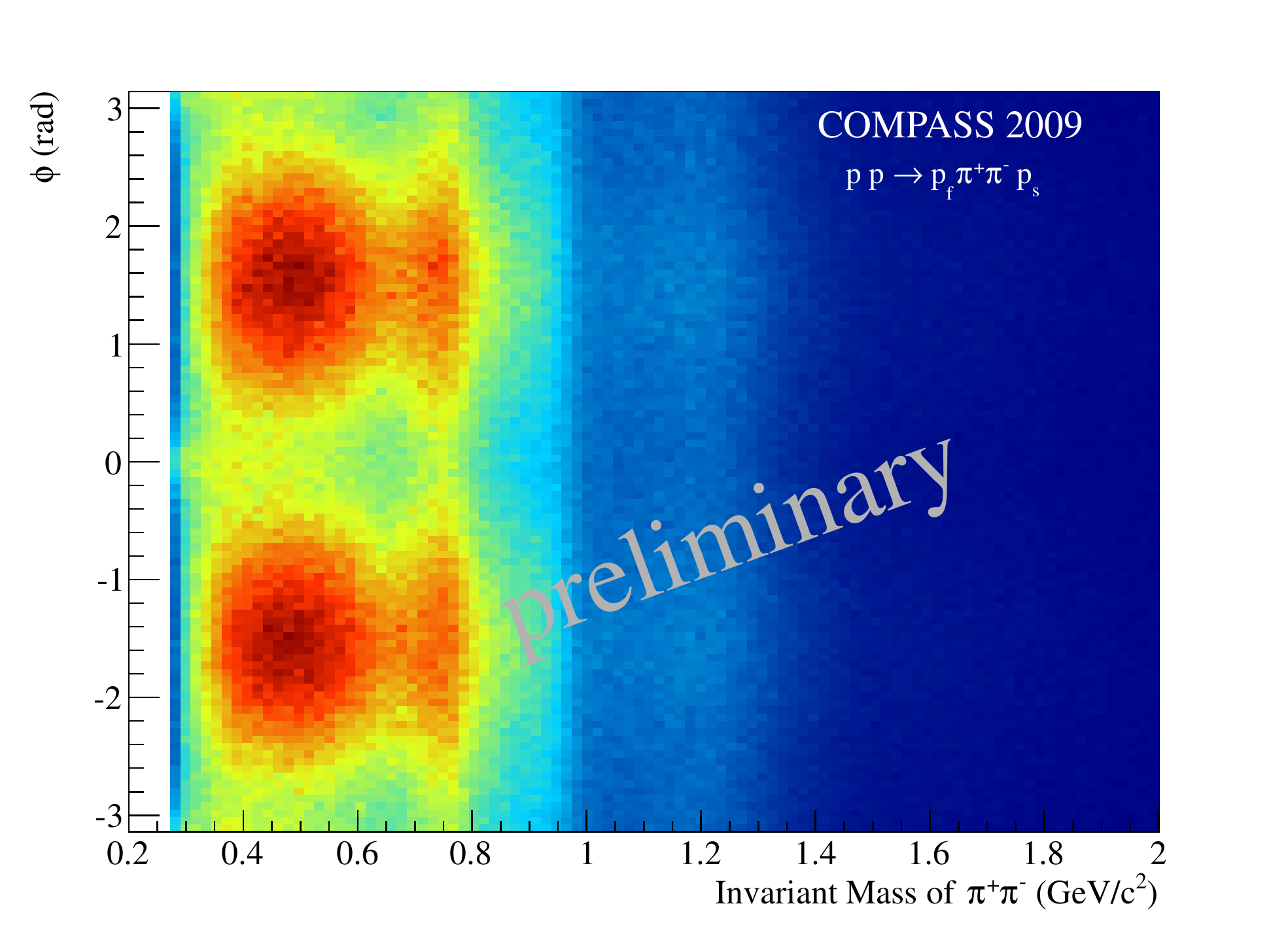}}
    \caption{\em  Decay variables as a function of the $\pi^+\pi^-$ mass}
    \label{fig:pwa}
  \end{figure}

  The $y$-axis of the right-handed coordinate system is defined by the cross product of the momentum vectors of the two exchange particles in the $pp$ centre-of-mass. Two variables specify the decay process, namely the polar and azimuthal angles ($\cos(\theta), \phi$) of the $\pi^-$ in the di-pion centre-of-mass frame relative to these axes.
  Figure~\ref{fig:pwa} illustrates the distribution of these decay variables as a function of the di-pion mass.

  The decay amplitudes for a two-pseudoscalar system with relative angular momentum $\ell$ and its projection on the quantisation axis $m$ is given by the spherical harmonics $Y^\ell_m(\theta, \phi)$ in spherical coordinates. To profit from the fact that the strong interaction conserves parity, a reflection operator was introduced in \cite{chu75} as a parity operator followed by a rotation around the $y$-axis such that all momenta relevant to the production process are recovered. Introducing a quantum number $\varepsilon$ which is called {\em reflectivity} and which can have the values $\pm 1$, the eigenstates of this reflection operator can be constructed as:
  \begin{equation}
    Y^{\varepsilon \ell}_m(\theta, \phi) \equiv c_m \left [ Y^\ell_m (\theta , \phi) - \varepsilon (-1)^m Y^\ell_{-m}(\theta, \phi) \right ]
    \label{eq:eigen}
  \end{equation}
  with a normalisation constant $c_m$. The two classes of states $\varepsilon = \pm1$ correspond to different production processes in the asymptotic limit~\cite{got64} and can consequently not interfere.

 With the complex {\em transition amplitudes} $T_{\varepsilon l m}$ and an explicit incoherent sum over the reflectivities, the intensity in mass bins can be expanded in terms of partial waves:
  \begin{equation}
    I(\theta, \phi) = \sum_\varepsilon \left \vert \sum_{\ell m} T_{\varepsilon \ell m}Y^{\varepsilon \ell}_m(\theta, \phi) \right \vert^2
  \end{equation}
  
  We adopt the notation $J^\varepsilon_m$ from \cite{bar99} to construct the basic wave set $\mathbf{S^-_0}, P^-_0, P^-_1, D^-_0, D^-_1$ and $\mathbf{P^+_1}, D^+_1$. Since the overall phase for each reflectivity is indeterminate, one wave in each class can be defined real, which means the imaginary part of that transition amplitude is fixed to zero. These so-called {\em anchor} waves are emphasised in bold font.

  An {\em extended likelihood} fit in $20\,\mathrm{MeV}/c^{2}$ mass bins is used to find the parameters $T_{\varepsilon l m}$, such that the acceptance corrected model $I(\theta, \phi)$ matches the measured data best.

  \section{Ambiguities}

As shown in \cite{sad91,chu97}, the angular dependency of the intensity $I(\Omega)$ for two-pseudoscalar final states can generally be expressed in terms of $|G(u)|^2$ by the introduction of a variable $u \equiv \tan(\theta/2)$. The function $G(u)$ is a complex polynomial of the order of $2\ell$, where $\ell$ is the the highest considered spin in the system. This polynomial can be factorised in terms of its complex roots  $u_k$, the so-called {\em Barrelet-zeros}~\cite{bar72}. Since the function $G(u)$ only enters as absolute square in the expression for the angular distribution, the complex conjugate of a root $u_k^*$ is an equally valid solution. That means, that there are in general $2^{2\ell-1}$ different ambiguous solutions which result in exactly the same angular distribution. This ambiguity has to be resolved by physical arguments.

\begin{figure}[t]
    \begin{center}
      \subfigure[]{
        \includegraphics[width=.45\textwidth]{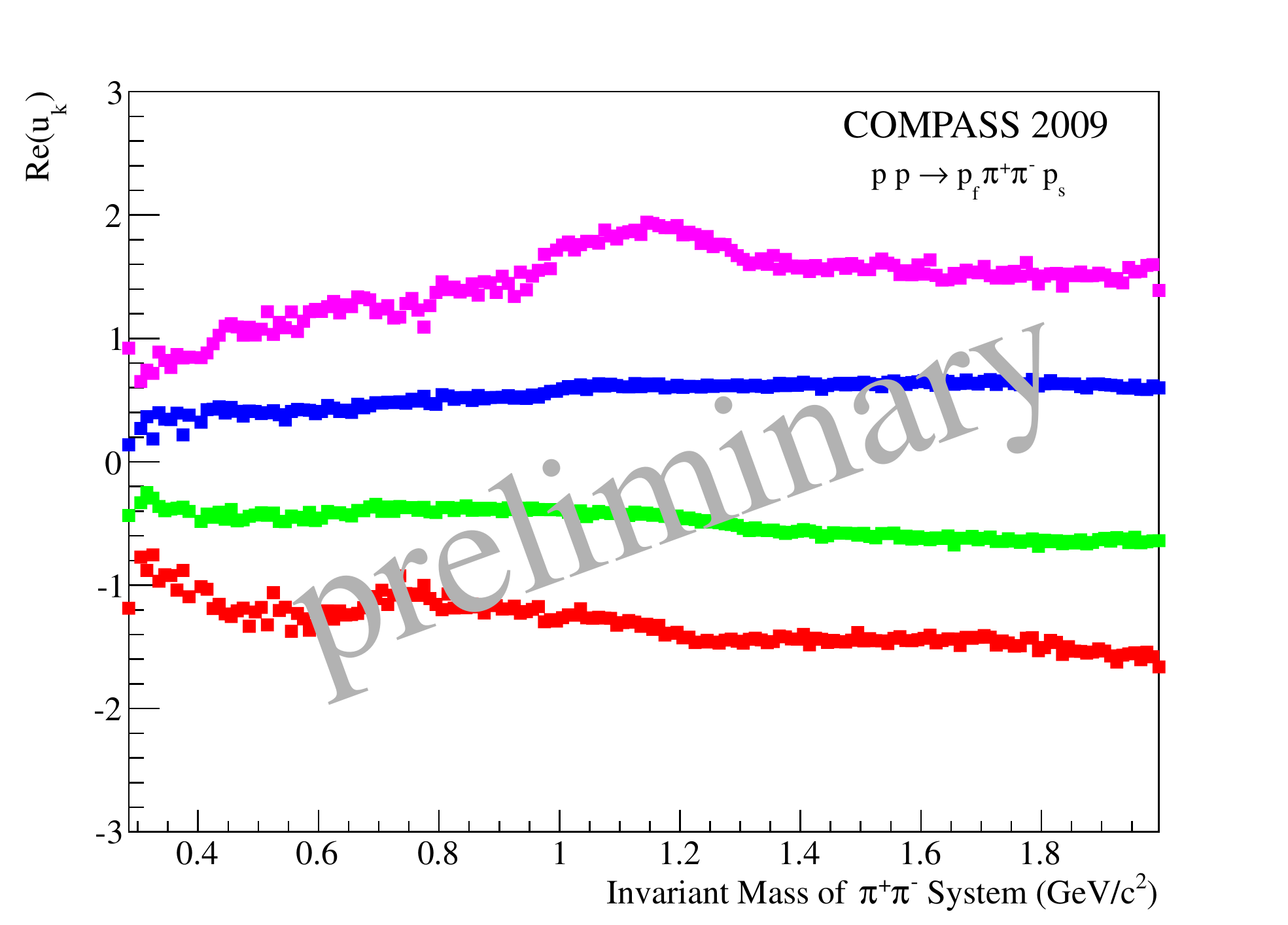}
      }
      \subfigure[]{
        \includegraphics[width=.45\textwidth]{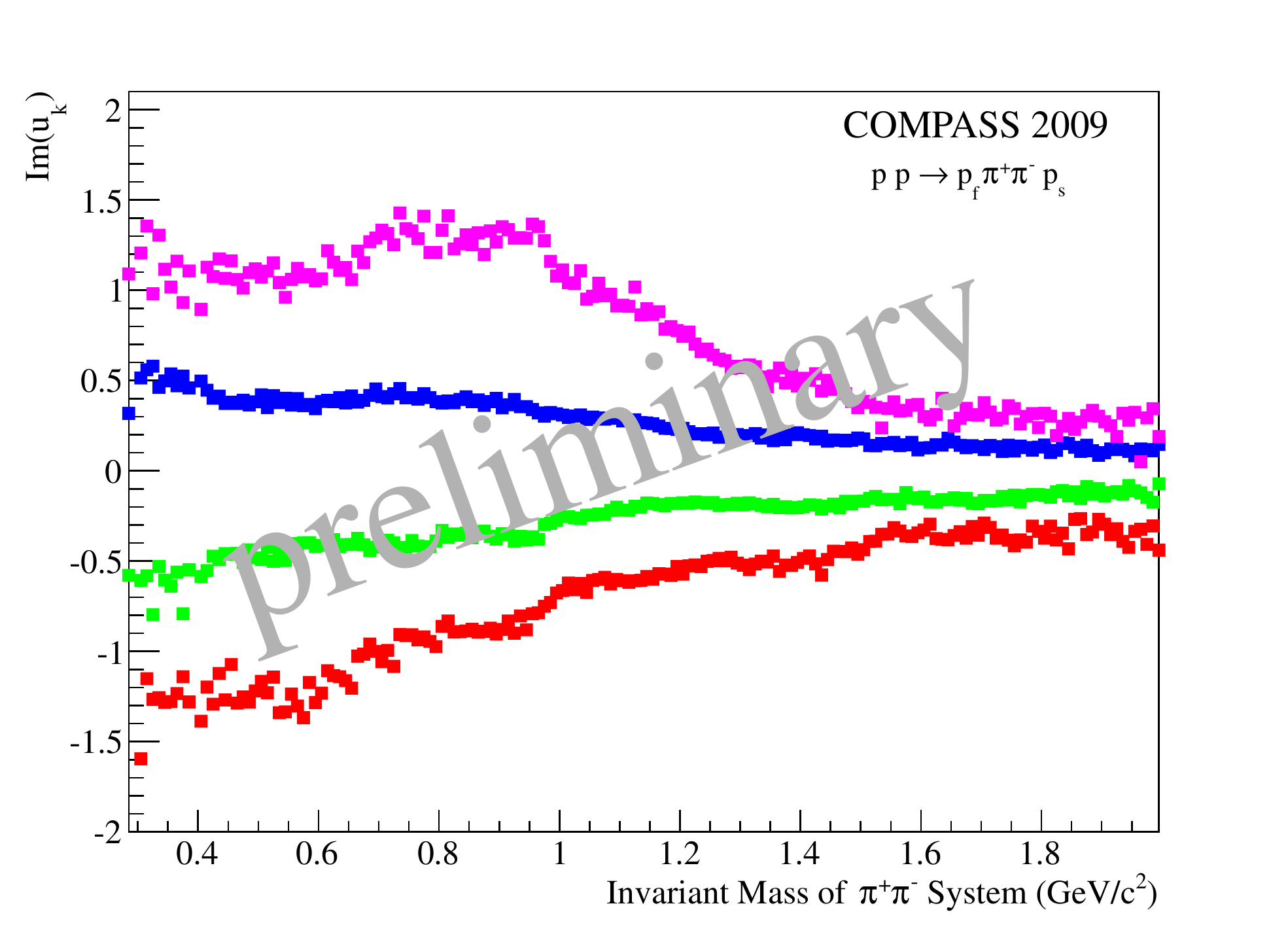}
      }
    \end{center}
    \caption[{\em Barrelet Zeros}]{{\em  The (a) real and (b) imaginary parts of the Barrelet-zeros as a function of the $\pi^+\pi^-$ mass}}
    \label{fig:roots}
  \end{figure}

The system of $S$, $P$ and $D$ waves used in the presented analysis has eight ambiguous solutions. By using different random starting values for the fit to the same data and MC sample, it was shown experimentally that very different solutions can be obtained. However, the fitted production amplitudes $T_{\varepsilon l m}$ for one single attempt can be used to calculate all eight solutions analytically~\cite{chu97}. In order to find the complex polynomial roots numerically, {\em Laguerre's method} was applied. Figure~\ref{fig:roots} illustrates the real and imaginary parts of these four roots for all mass bins, where a sorting depending on the real part has been performed. They are well separated from each other and do not require a special linking procedure. The imaginary parts do not cross the real axis, hence bifurcation of the solutions does not pose a problem and the solutions can be uniquely identified. 

  By fixing $u_1$ and allowing $u_{2,3,4}$ to undergo complex conjugation, the entire set of eight ambiguous solutions is calculated. For six solutions, most of the intensity is formed by one single wave. This is clearly unphysical, since we know of at least one resonance in each of the spins $0$, $1$ and $2$. The choice among the remaining solutions is not evident, the intensity and phase distributions are very similar.

  \begin{figure}[ht!]
    \begin{center}
      \includegraphics[width=.93\textwidth]{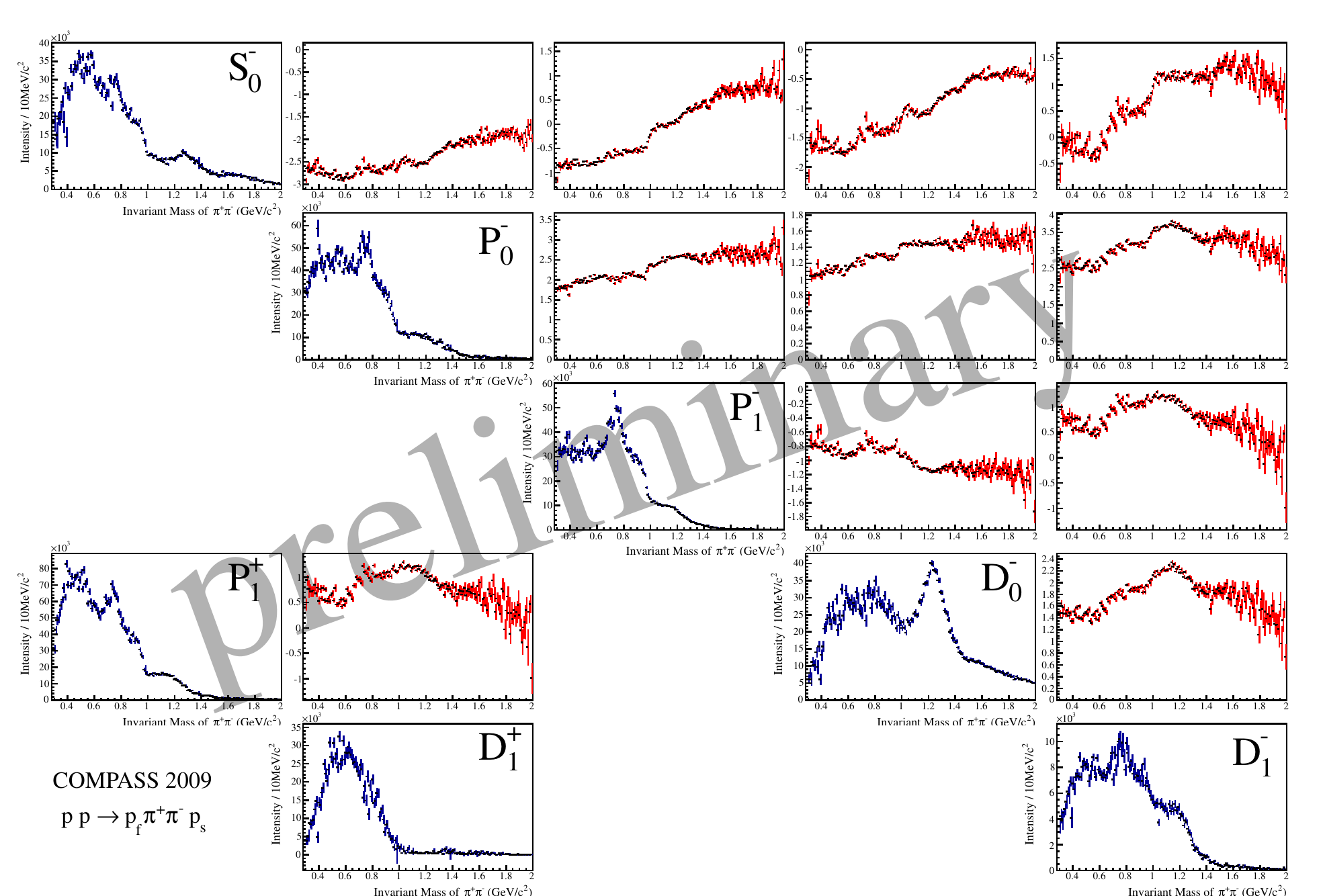}
    \end{center}
    \caption[{\em Physical Solution}]{\em Solution compatible with physical constraints, intensities (blue) and relative phases (red)}
    \label{fig:phys}
  \end{figure}

  As a last step, the calculated values for one chosen solution are reintroduced to the fit as starting values, which probes the convergence and provides the correct covariance matrix. Figure~\ref{fig:phys} shows the intensities and phases for one solution compatible with the physical constraints, depicted in a matrix form. A clear signal from $f_2$(1270) can be seen in the $D^-_0$ wave, accompanied by a phase motion with respect to the $S^-_0$. There is also evidence for $\rho$(770) in both $P^-_0$ and $P^-_1$. At masses well below $1\,\mathrm{GeV}/c^2$, however, the scalar wave does not show the expected threshold enhancement.

  To evaluate the fit quality, the decay amplitudes %$Y^{\varepsilon \ell}_m$
 of phase-space Monte-Carlo events are weighted by the production amplitudes $T_i$ obtained in the minimisation. The projected angular distributions show very good agreement with the data.
 
  \section{Outlook}
  \label{sec:DaO}

  With these proceedings, we want to show that we are able to select a centrally produced sample and describe the main features of the data in terms of partial waves. Using the methods from~\cite{bar99}, we can reproduce the analysis with comparable results.

  The complicated problem of central production was approached following a model employed by WA102, an earlier experiment at CERN, inheriting some limitations and assumptions. In the meantime, other ideas were developed based on their data and analyses which can help to determine spin and parity of particles produced in central exclusive processes~\cite{kai03}. In addition to measuring the decay products, it can have particular advantages to study angular correlations between the outgoing protons. In the long run, we might study the data with a combination of both.

  In order to obtain a complete picture of the light scalar meson sector, the decay of identified resonances into different final states is needed to understand their composition. Since the trigger was not selecting a particular final state and the generalisation of the analysis of the two-pion system to other two-pseudoscalar final states is trivial, the next natural step will be the analysis of the $K^+K^-$ system. Other available channels are $K_sK_s$, $\pi^0\pi^0$ and $\eta\eta$. We expect that COMPASS will be able to shed light on this sector, outperforming previous experiments in terms of data as well as resolution.

\end{document}